\documentstyle[aps,prl,multicol,epsf,epsfig]{revtex}

\begin{document}
\title{Interaction between Kondo impurities in a quantum corral}
\author{G. Chiappe$^{a}$ and A. A. Aligia$^{b}$}
\address{$^{a}$ Departamento de F\'{\i }sica, FCEyN Universidad de Buenos Aires,\\
Pabell\'{o}n I, Ciudad Universitaria, (1428) Buenos Aires, Argentina.\\
$^{b}$ Centro At\'{o}mico Bariloche and Instituto Balseiro, Comisi\'{o}n
Nacional de Energ\'{\i }a At\'{o}mica, 8400 Bariloche, Argentina.}
\date{Received \today }
\maketitle

\begin{abstract}
We calculate the spectral densities for two impurities inside an elliptical
quantum corral, using exact diagonalization in the relevant Hilbert subspace
and embedding into the rest of the system. Fore one impurity, the space and
energy dependence of the change in differential conductance $\Delta dI/dV$
observed in the quantum mirage experiment is reproduced. In presence of
another impurity, $\Delta dI/dV$ is very sensitive to the hybridization
between impurity and bulk. The impurities are correlated ferromagnetically
between them. A hopping $\gtrsim 0.15$ eV between impurities destroys the
Kondo resonance.
\end{abstract}



\twocolumn[
\hsize\textwidth\columnwidth\hsize\csname@twocolumnfalse\endcsname
]
\narrowtext
In recent years, the manipulation of single atoms on top of a surface using
scanning tunneling microscopy (STM) was made possible,\cite{eig} and quantum
corrals have been assembled by depositing a closed line of atoms or
molecules on noble metal surfaces .\cite{cro,hel,man} The local conduction
spectral density of states $\rho _{c}(r,\omega )$, measured by differential
conductance $dI/dV$ reveals patterns that remind the wave functions of
two-dimensional noninteracting electrons under the corresponding confinement
potential. In a recent experiment, a Co atom has been placed at a focus of
an elliptic quantum corral, and the corresponding Kondo feature is
observed
not only at that position, but also at the other focus, where a ``mirage''
is formed as a consequence of the quantum interference.\cite{man} Several
variants of this experiment, some of them involving several impurities (Co
atoms) and eventually mirages 
inside the corral, are being performed.\cite{man2} The main features
of the observed space and voltage dependence of $dI/dV$ have been reproduced
by several theories.\cite{aga,fie,por,wei,ali,ali2} Since in Refs. \cite
{aga,fie,por} the density of states per spin at the impurity $\rho
_{d}(\omega )$ is assumed rather than calculated, these theories cannot
account for the interaction between impurities. In Ref. \cite{wei} the Kondo
effect is absent, and perturbation theory in the Coulomb repulsion $U$ \cite
{ali,ali2} is restricted to small values of $U$.

The aim of the present work is to present a theory of the quantum mirage
which is able to reproduce the experimental results for the case of one
impurity and give reliable predictions when more than one impurity is inside
the corral. We show that experiments with two impurities can elucidate the
role of the direct hybridization between the impurity and the bulk $V_{b}$.
Scattering theories \cite{hel,aga,fie} obtained an excellent agreement with
experiment assuming that the resonant level width due to hybridization with
bulk states $\delta _{b}$ is as large as that due to the surface $\delta
_{s} $. On the other hand, the larger density of $s$ and $p$ states at the
surface \cite{eu} and the rapid decay of the hybridization matrix elements
with distance suggest that $\delta _{b}$ is negligible, and the experiment
for one impurity can also be explained if $V_{b}=0$.\cite{ali,ali2} A calculation
of $\delta _{b}$ has not been made and experimentally the situation is still
unclear. 
The role of $V_{b}$ is not only crucial for a correct theory of the mirage experiment, 
but also for the general understanding of the interaction between metallic
surfaces and adsorbates.
Since actually $\delta _{b}$ was introduced as a phenomenological
parameter which takes into account the electrons lost in the scattering
process,\cite{hel,aga,fie} one expects that if $\delta _{b}=\delta _{s}$,
the interaction between impurities is roughly a fourth of that for $V_{b}=0$
if the same total width $\delta _{b}+\delta _{s}$ is kept.

We obtain the ground state of the Anderson model in a cluster which contains
one or two impurities and the relevant conduction states inside a hard wall
ellipse using the Lanczos method. These states are then mixed with bulk
states using an embedding method.\cite{fer} This embedding is essential to
describe the low energy physics.\cite{hal} The average separation between
the relevant conduction states $d\sim 100$ meV \cite{por,ali} is much larger
than the Kondo temperature $T_{K}\sim 5$ meV.\cite{man} Under these
circumstances a Kondo peak at the Fermi level $\epsilon _{F}$ is absent in
the finite system \cite{ali,thi} and the experimental line shape for $dI/dV$
cannot be reproduced (unless an artificial Lorentzian broadening is
introduced). This is confirmed by our calculations.

The Hamiltonian can be written as: 
\begin{eqnarray}
H &=&\sum_{j\sigma }\varepsilon _{j}c_{j\sigma }^{\dagger }c_{j\sigma
}+E_{d}\sum_{i\sigma }d_{i\sigma }^{\dagger }d_{i\sigma
}+U\sum_{i}d_{i\uparrow }^{\dagger }d_{i\uparrow }d_{i\downarrow }^{\dagger
}d_{i\downarrow }  \nonumber \\
&&+\sum_{ij\sigma }V[\varphi _{j}(R_{i})d_{i\sigma }^{\dagger }c_{j\sigma }+%
\text{H.c.}]+H^{\prime }.  \label{ham}
\end{eqnarray}
Here $c_{j\sigma }^{\dagger }$ creates an electron on the $j^{th}$
conduction eigenstate of a hard wall elliptic corral with wave function $%
\varphi _{j}(r)$ \cite{note} and $d_{i\sigma }^{\dagger }$ is the
corresponding operator for the impurity at site $R_{i}$. The hybridization
of these states with bulk states of the same symmetry is described by $%
H^{\prime }$. We assume that each of the impurity and conduction states
mixes with a different continuum of bulk states:

\begin{equation}
H^{\prime }\cong t\sum_{j\sigma }(c_{j\sigma }^{\dagger }b_{j\sigma }+\text{%
H.c.})+V_{b}\sum_{i\sigma }(d_{i\sigma }^{\dagger }b_{i\sigma }+\text{H.c.}).
\label{h2}
\end{equation}
The $b_{l\sigma }$ represent bulk states for which the unperturbed density
is 0.05 states/eV, similar to the density of bulk $s$ and $p$ states.\cite
{eu} Approximation (\ref{h2}) is justified by comparison of the 
non-interacting Green functions for hard wall corrals and more realistic
boundary potentials.\cite{ali2,cor}

The dressed matrix ${\bf G}$ describing the one-particle Green function is
calculated by solving the Dyson equation ${\bf G=g+g}H^{\prime }{\bf G}$,
where ${\bf g}$ is the corresponding matrix for $H^{\prime }=0$.\cite{fer}
This equation is exact for $U=0$ or $H^{\prime }=0$, and in the general case
represents an infinite sum of particular diagrams in perturbation theory in $%
H^{\prime }$ (the chain approximation\cite{cha}). The $\varphi _{j}(r)$ are
obtained as described elsewhere.\cite{ali} We choose the ellipse with
eccentricity $e=1/2$ and size such that the state $j=42$ lies at $\epsilon
_{F}$.\cite{man} The change in $dI/dV$ ($\Delta dI/dV$) after an impurity is
placed inside the corral is determined by the conduction states which lie
near $\epsilon _{F}$ and have a strong amplitude $\left| \varphi
_{j}(R_{i})\right| $ at the impurity position. For $R_{i}$ at one focus they
are $j=32$, 35, 42 and 51.\cite{ali} We have also included $j=24$ and 62,
although this inclusion leads to negligible changes in the results. We took
the impurity parameters $E_{d}=-1$ eV and $U=3$ eV.\cite{ujs} We consider
first the case $V_{b}=0$ and one impurity at the left focus ($R_{i}=(-0.5a,0)$). 
The value $V=0.04$ eV was chosen to lead to the observed width of $\Delta
dI/dV$. The remaining parameter $t$ controls the amplitude of the mirage at
the right focus.

In Fig. 1(a) we represent the resulting impurity spectral density $\rho
_{d}(\omega )$ for two values of $t$. A clear Kondo peak is obtained and for 
$t\gtrsim 0.3$ eV its width is very weakly dependent on $t$. Instead, for $%
t\rightarrow 0$, the peak splits into two very narrow peaks out of $\epsilon
_{F}$. In contrast to $\rho _{d}(\omega )$, the magnitude of the change in
the conduction density $\Delta \rho _{c}(r,\omega )$ at the empty focus ($%
r=-R_{i}$) is quite sensitive to $t\gtrsim 0.3$ eV: as $t$ increases, the
width of the conduction states increases, the weight of the states 32, 35
and 51 (odd under the reflection through the minor axis of the ellipse $%
\sigma $) at $\epsilon _{F}$ increases, and the depression of $\rho
_{c}(-R_{i},\omega )$ decreases as a consequence of the negative
interference of these states with the even state 42.\cite{ali} The
differential conductance $dI/dV$ at zero temperature is proportional to the
density $\rho _{f}$ of the state\cite{schi}

\begin{equation}
f_{\sigma }(r)=\sum_{j}\varphi _{j}(r)c_{j\sigma }+qd_{j\sigma }.  \label{q}
\end{equation}
$q$ is related to Fano`s interference parameter and represents the effect of
a direct tunneling from the tip to the impurity. Therefore, it is relevant
only very near the impurity. For $q=0$, $\rho _{f}(r,\omega )=\rho
_{c}(r,\omega )$. In Fig. 1 (b) we represent the effect of adding the
impurity on $\rho _{f}(\pm R_{i},\omega )$ ($\Delta \rho _{f}\sim \Delta
dI/dV$). At the impurity site $R_{i}$, $\Delta \rho _{c}(r,\omega )$ is
asymmetric and smaller at the right of the valley. This is a consequence of
the asymmetry of the hybridization around $\epsilon _{F}$ ($\left| \varphi
_{51}(R_{i})\right| >\left| \varphi _{35}(R_{i})\right| $). A symmetric line
shape, as observed in the experimental $\Delta dI/dV$ is restored for $q\sim
1$. The effect of this $q$ is consistent with the fact that on a clean
surface, $\Delta dI/dV$ is {\em larger} at the right of the peak. Another
nice fact is that the minimum of $\Delta \rho _{f}$ for $q=1$ lies at the
experimental position 1 meV. At the right focus ($r=-R_{i}$) we obtain a
similar valley, although slightly asymmetric and shifted to the left.
Increasing $t$ from 0.4 to 0.5, the magnitude of this valley is strongly
reduced (its minimum is shifted above -5/eV) but its shape and width is
retained. At the impurity position there are no significant changes.

The space dependence of $\Delta \rho _{f}$ for $q=0$ is represented in Fig.
2. As in the experimental $\Delta dI/dV$, the main features of $\left|
\varphi _{42}(r)\right| ^{2}$, attenuated at the right focus, are displayed.
Thus, the theory reproduces the space and energy dependence of $\Delta dI/dV$
observed in the experiment.\cite{man} All results so far agree 
semiquantitatively
with perturbative calculations.\cite{ali,ali2} To see how the results
change if $\delta _{b}\cong \delta _{s}$ is assumed, we have reduced $V$ by
a factor $\sqrt{2}$. This should reduce $\delta _{s}$ by a factor 2.
Increasing $V_{b}$ from zero to 1.2 eV, the original width of $\rho _{d}$ is
restored. The intensity is reduced by a factor $\sim 2$ (due to the strong
energy dependence of $\delta _{s}$). $\Delta \rho _{c}$ turns out to be $%
\sim 4$ times smaller. The additional factor 2 can be understood from the
fact that the change in conduction electron Green function is proportional
to $V^{2}G_{d}(\omega )$, where $G_{d}(\omega )$ is the impurity Green
function.\cite{ali} Except for these factors, the results are surprisingly
similar to the previous ones. Some of them will be displayed in Fig. 4.

We now turn to the case of two impurities, one at each focus, for $V_{b}=0$.
The spectral density for one of these impurities is represented in Fig. 3
(a). Comparison with the previous case (Fig. 1), shows that the peak around $%
\epsilon _{F}$ broadens (by a factor $\sim 1.5$), looses intensity and
shifts to lower energies. In addition, another very narrow peak appears $%
\sim 13$ meV below $\epsilon _{F}$. An analysis of the energy dependence of
the density of the individual conduction states shows that the broad peak
around $\epsilon _{F}$ is due to hybridization with even states (mainly 42),
while the narrow peak reflects the hybridization of the impurity states with
odd states (mainly 51). The difference in $dI/dV$ with respect to the empty
corral is however, not so different as in the previous case. This is due to
the effect of the unperturbed Green functions of the conduction states and
is also present in the one impurity case.\cite{ali} Nevertheless, a decrease
in the amplitude and a broadening of the depression should be observed in $%
\Delta dI/dV$ and seems in qualitative agreement with recent experiments.%
\cite{man2} The space dependence is similar to that for one impurity (Fig.
2) but it is of course, symmetric under reflection through the minor axis $%
\sigma $, and not attenuated at the right focus.

Qualitatively, the shape of $\rho _{d}$ can be understood looking at the
non-interacting case $U=0$, $E_{d}\sim $ $\epsilon _{F}$. In this case, for
one impurity, the Kondo peak is replaced by a Lorentzian near $\epsilon _{F}$%
. For two impurities, a change of basis of the $d$ orbitals to $e_{\sigma
}=(d_{1\sigma }+d_{2\sigma })/\sqrt{2}$, $o_{\sigma }=(d_{1\sigma
}-d_{2\sigma })/\sqrt{2}$, separates the problem into those corresponding to
even and odd states under $\sigma $. The even state hybridizes mainly with
conduction state 42, to form a resonance near $\epsilon _{F}$, roughly twice
wider than for one impurity due to the larger effective hybridization.
Instead, the odd state $o_{\sigma }$ is displaced towards lower energies due
to hybridization with state 51. The interactions should modify the
quantitative details of this picture. However, we expect that it remains
qualitatively valid, as suggested by the above mentioned energy distribution
of the different conduction states.

To gain insight into the nature of the ground state, we have also calculated
spin-spin correlation functions for $t=0$. A reliable method to include $%
H^{\prime }$ in these calculations has not been developed yet. For one
impurity we obtain $\langle {\bf S}_{i}\cdot {\bf s}_{42}\rangle =-0.73$,
where ${\bf S}_{i}$ is the spin of the impurity $i$ and ${\bf s}_{j}$ is the
spin of the conduction state $j$. This value is close to the minimum
possible one -3/4.  For $j\neq 42$, $\langle {\bf S}_{i}\cdot {\bf s}%
_{j}\rangle $  are very small, but this, and the large magnitude of $\langle 
{\bf S}_{i}\cdot {\bf s}_{42}\rangle $, are affected to a certain degree by
the neglect of $H^{\prime }$ in this calculation. The space dependence of $%
\langle {\bf S}_{i}\cdot {\bf s}(r)\rangle $, where  ${\bf s}(r)$ the
conduction spin at position $r$ follows closely $\left| \varphi
_{42}(r)\right| ^{2}$. For two impurities we find $\langle {\bf S}_{i}\cdot 
{\bf s}_{42}\rangle =-0.47$ and$\ \langle {\bf S}_{1}\cdot {\bf S}%
_{2}\rangle =0.21$. In the limit of large $U$, one expects that the main
features of the spin dynamics for $V_{b}=0$ are described by the Hamiltonian 
$H_{0}=J({\bf S}_{1}+{\bf S}_{2})\cdot {\bf s}_{42}$, where $J>0$ is the
Kondo coupling. The ground state of this Hamiltonian is a doublet  in which
the impurity spins are correlated ferromagnetically between them ($\langle 
{\bf S}_{1}\cdot {\bf S}_{2}\rangle =1/4$) and antiferromagnetically with
state 42 ($\langle {\bf S}_{i}\cdot {\bf s}_{42}\rangle =-1/2$). These
values are near to those we find. The effect of the hybridization of
state 42 with bulk states can be modelled by a tight binding Hamiltonian in
terms of Wilson's orbitals. $H_{0}$ is the strong coupling fixed point of
Wilson's renormalization group. An analysis of the stability of this fixed
point using perturbation theory as in Ref. \cite{allub} leads to the
conclusion that the ground state is a doublet for $V_{b}=0$. However, we
expect that as soon as $V_{b}\neq 0$, the doublet is screened at a very low
temperature.

For the set of parameters corresponding to $\delta _{b}\cong \delta _{s}$, $%
\rho _{d}(\omega )$ is much more similar to the one impurity case, although
a structure reminiscent of a splitting is also present near its maximum. In
contrast to the case of $V_{b}=0$, when a second impurity is added, the
depression in $\Delta dI/dV$ at one impurity site $R_{i}$ increases and its
width is roughly the same (see Fig. 4). Comparison with results when $t$ is
increased from 0.4 eV to $t=0.5$ eV (not shown) suggests that when $\delta
_{b}\cong \delta _{s}$, $\Delta dI/dV$ at $\pm $ $R_{i}$ for two impurities
is roughly the sum of the results at $R_{i}$ and $-R_{i}$ for one impurity.
This is what one would expect if the interaction is very small.

Coming back to the case $V_{b}=0$, we have also verified that qualitatively
similar features in $\Delta dI/dV$ are obtained at one focus, if one
impurity is placed there and the second impurity is put at another extremum
of $\varphi _{42}(r)$, like (0.22$a$,0) (instead of placing it at the other
focus). In this case, the spectral densities at (0.22$a$,0) have some
additional structure due to an important admixture of the state 41.\cite
{note2} In contrast, if both impurities are placed close to the same focus
and near each other, a moderate hopping $t\sim 0.15$ eV or larger between
them is sufficient to destroy the Kondo resonance. In particular $\Delta
dI/dV$ becomes flat and featureless near $\epsilon _{F}$.

In summary, we have studied the spectral density for impurities inside a
quantum corral, using a many-body approach which treats exactly the
correlations in the impurities and their hybridization with the relevant
conduction states at the surface, and treats approximately the hybridization
with bulk states. We have been able to reproduce the main features of the
mirage experiment for one impurity inside the corral. The experiment for one
impurity cannot determine the relative importance of the direct
hybridization of the impurity with bulk states, unless the tunneling matrix
elements and other details are known accurately. Instead, for two impurities
inside the corral, the differential conductance is very sensitive to this
hybridization. For the parameters of the experiment, the spins of both
impurities are antiferromagnetically coupled with the conduction electrons,
and ferromagnetically correlated between them provided they are placed
sufficiently far apart, so that the hopping between them can be neglected.
If this hopping is larger than 0.15 eV, there is a tendency to form a
singlet state between both impurity spins and the Kondo resonance disappears.
To our knowledge, this is the first theory which is able to describe the 
line shape of the differential conductance when more than one Kondo impurity 
is inside the quantum corral. 

This work benefitted from PICT 03-00121-02153 of ANPCyT and PIP 4952/96 of
CONICET. We are partially supported by CONICET.

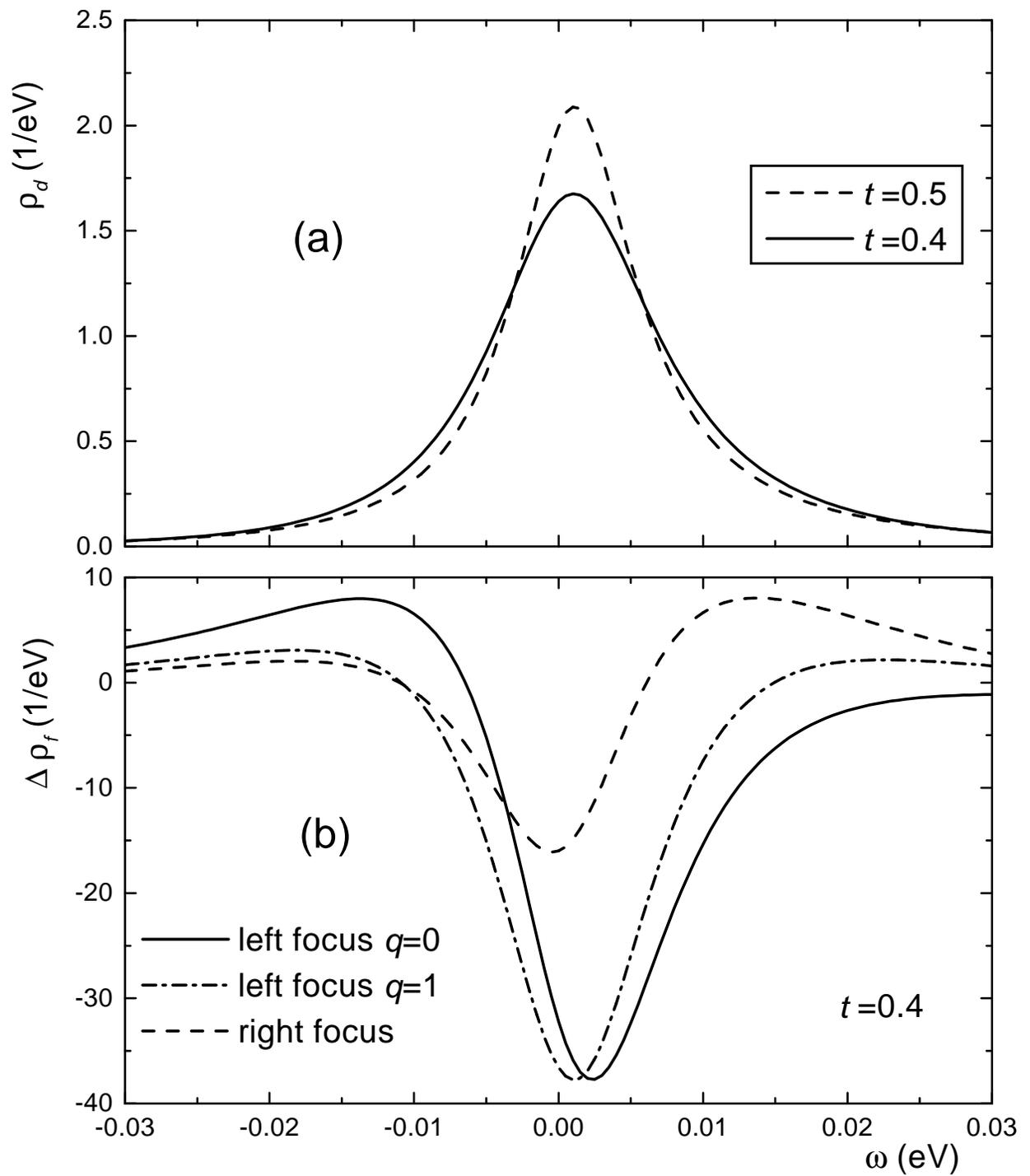
\figure{\noindent Fig. 1: (a) Impurity spectral density as a function of
energy for two values of }$t${. (b) Change in the density of the mixed state 
}$f_{\sigma }$ (Eq.(\ref{q})) at the impurity site (left focus) for two
values of $q$ and at the other focus for $t=0.4$ eV{.}

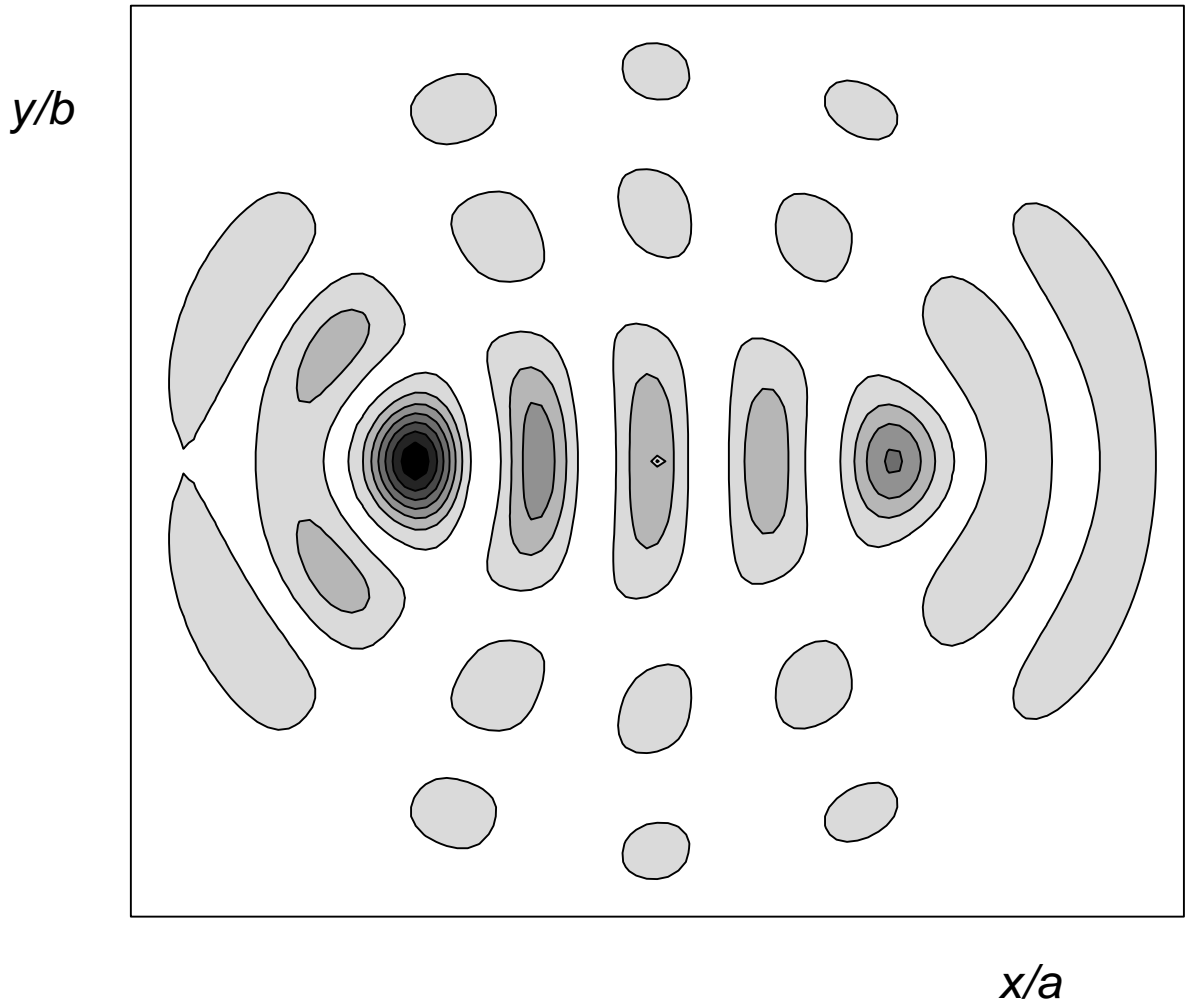
\figure{\noindent Fig 2: Contour plot of $\Delta \rho _{c}(r,\omega )$ for  
$t=0.4$ eV and {\ $\omega =10$ meV.}

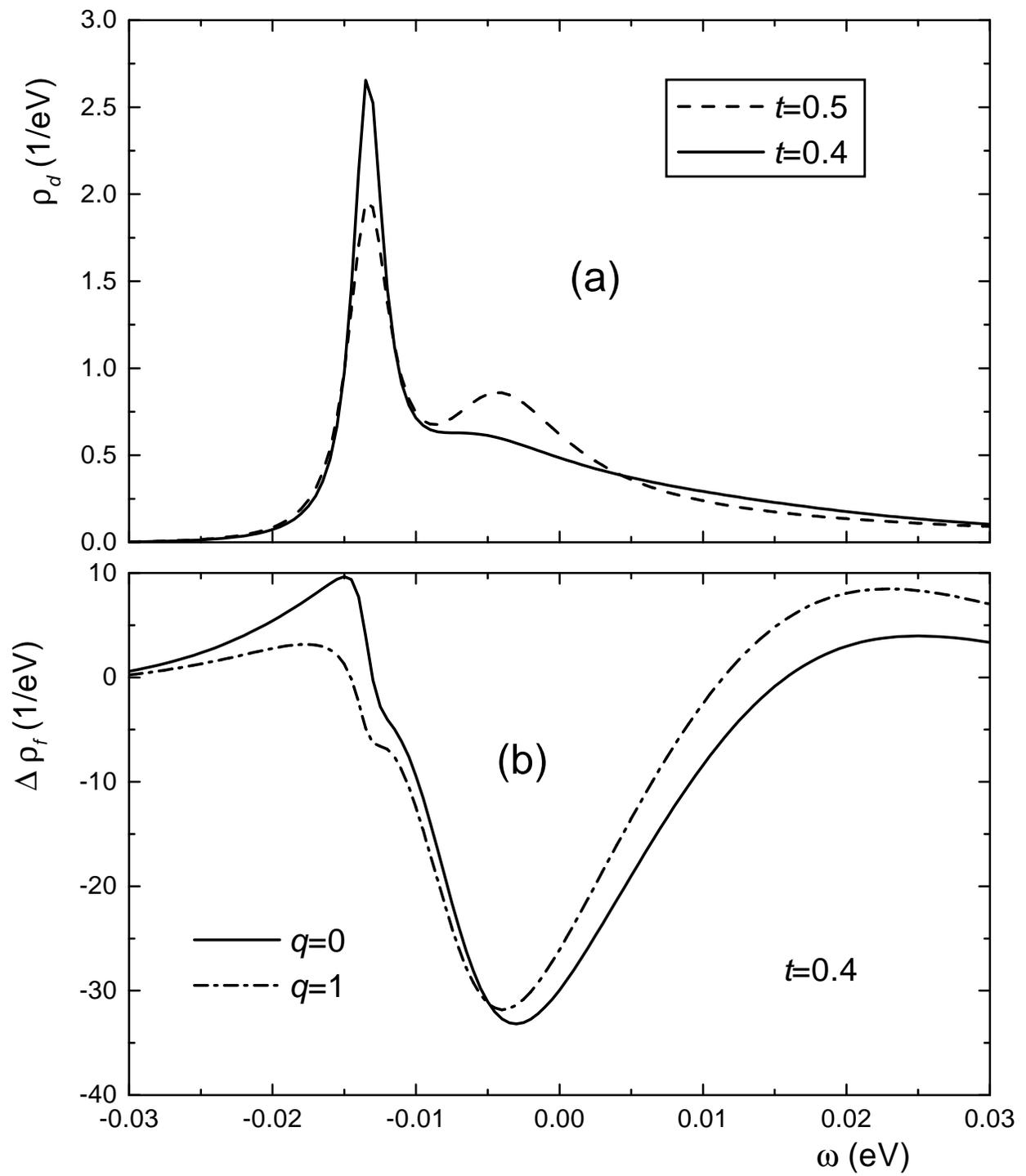
\figure{\noindent Fig 3: (a) Impurity spectral density for one impurity at
each focus and two values of }$t${. (b) Change in the density of the mixed
state }$f_{\sigma }$ after addition of both impurities (Eq.(\ref{q})), at
one impurity site for two values of $q$. Parameters are $V=0.04$eV, $V_{b}=0$
and $t=0.4$ eV{.}

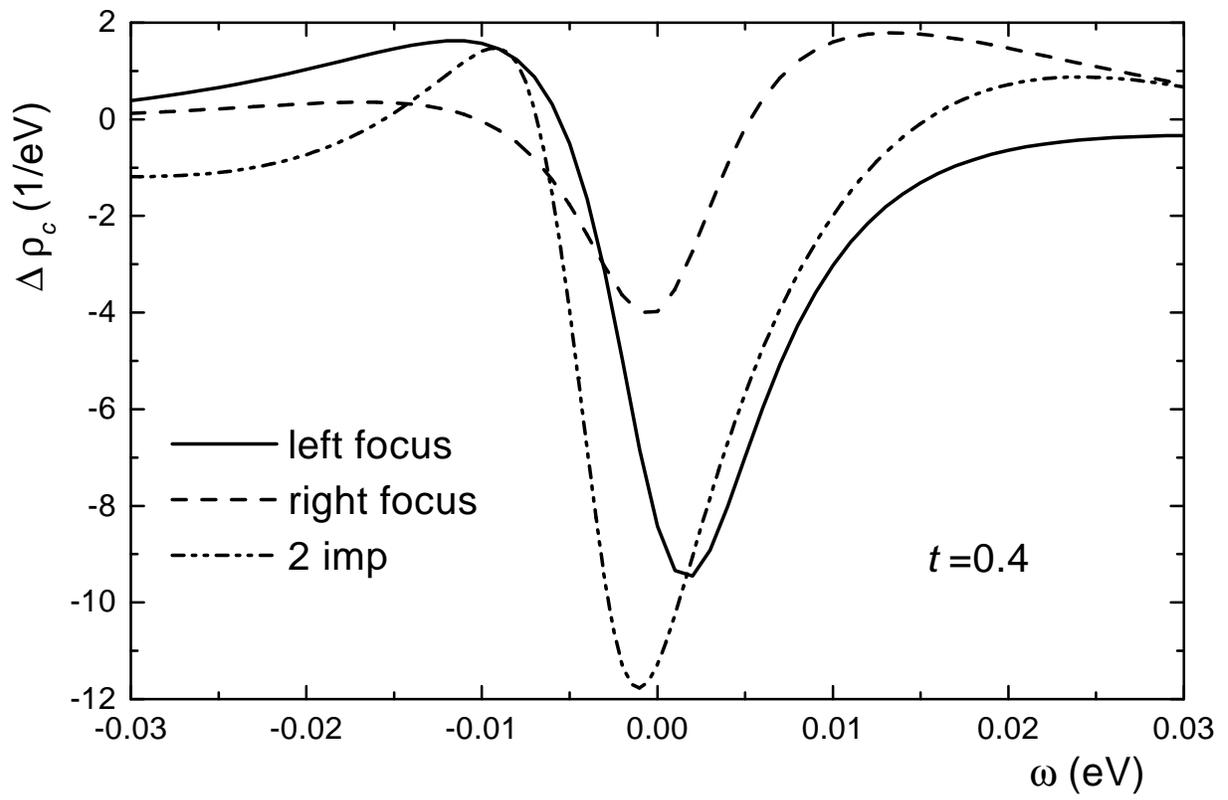
\figure{\noindent Fig 4: $\Delta \rho _{c}(r,\omega )$ as a function of 
$\omega $ for the case of one impurity at the left focus (full and dashed
lines) or one impurity at each focus (dashed dot dot line). Parameters are 
$V=0.04$eV/$\sqrt{2}$, $V_{b}=1.2$ eV and $t=0.4$ eV.}


\begin{references}
\bibitem{eig}  D.M. Eigler and E.K. Schweizer, Nature{\ {\bf 344}, 524
(1990).}

\bibitem{cro}  M.F. Crommie, C.P. Lutz, and D.M. Eigler, Science {\bf 262}{,
218 (1993).}

\bibitem{hel}  E.J. Heller {\it et al.}, Nature{\ {\bf 363}, 464 (1994).}

\bibitem{man}  H.C. Manoharan, C.P. Lutz, and D.M. Eigler, 
Nature{\ {\bf 403}, 512 (2000).}

\bibitem{man2}  H.C. Manoharan, PASI Conference, {\it Physics and Technology
at the Nanometer Scale} (Costa Rica, June 24 - July 3, 2001).

\bibitem{aga}  O. Agam and A. Schiller, Phys. Rev. Lett. {\bf 86}, 484
(2001).

\bibitem{fie}  G.A. Fiete {\it et al.}, Phys. Rev. Lett. {\bf 86}, 2392
(2001).

\bibitem{por}  D. Porras, J. Fern\'{a}ndez-Rossier, and C. Tejedor, Phys.
Rev. B {\bf 63}, 155406 (2001).

\bibitem{wei}  M. Weissmann and H. Bonadeo, Physica E {\bf 10}, 44 (2001).

\bibitem{ali}  A.A. Aligia, Phys. Rev. B {\bf 64}, 121102(R) (2001).

\bibitem{ali2}  A.A. Aligia, cond-mat/0110081.

\bibitem{eu}  A. Euceda, D.M. Bylander, and L. Kleinman, 
Phys. Rev. B {\bf 28}, 528 (1983).

\bibitem{fer}  V. Ferrari {\it et al.}, Phys. Rev. Lett. {\bf 82}, 5088
(1999); C.A. B\"{u}sser {\it et al.}, Phys. Rev. B {\bf 62,} 9907 (2000).

\bibitem{hal}  A numerical diagonalization without embedding has been
performed to study mirages on a spherical surface [K. Hallberg, A. Correa,
and C.A. Balseiro, cond-mat/0106082].

\bibitem{thi}  W.B. Thimm, J. Kroha, and J. von Delft, Phys. Rev. Lett. {\bf %
82}, 2143 (1999).

\bibitem{note}  Here the wave functions are adimensional and normalized as 
$\int dxdy\bar{\varphi}_{i}(r)\varphi _{j}(r)/(ab)=\delta _{ij}$, where 
$a$ ($b$) is the semimajor (semiminor) axis of the ellipse.

\bibitem{cor} A. Correa, K. Hallberg, and C.A. Balseiro, unpublished. 

\bibitem{cha}  E.V. Anda, J. Phys. C {\bf 14}, L1037 (1981); W. Metzner,
Phys. Rev. B {\bf 43}, 8549 (1991)..

\bibitem{ujs}  O.\'{U}js\'{a}ghy {\it et al.}, Phys. Rev. Lett. {\bf 85},
2557 (2000).

\bibitem{schi}  A. Schiller and S. Hershfield, 
Phys. Rev. B {\bf 61}, 9036 (2000).

\bibitem{allub}  R. Allub and A.A. Aligia, Phys. Rev. B {\bf 52}, 7987
(1995).

\bibitem{note2}  This calculation was made replacing the states 24 and 62,
which practically do not change the results, with states 41 and 49, which
have an important hybridization with the impurity at (0.22$a$,0).
\end{references}
\end{document}